\documentclass[12pt]{article}
\usepackage{amsmath,amssymb,amsfonts,graphicx}
\def\slash#1{\setbox0=\hbox{$#1$}#1\hskip-\wd0\hbox to\wd0{\hss\sl/\/\hss}}
\usepackage{epsf, cite}
\usepackage{epsfig}
\usepackage[all]{xy}
\usepackage{latexsym}

\makeatletter
\renewcommand\section{\@startsection {section}{1}{\z@}%
                                   {-3.5ex \@plus -1ex \@minus -.2ex}
                                   {2.3ex \@plus.2ex}%
                                   {\normalfont\large\bfseries}}
\renewcommand\subsection{\@startsection{subsection}{2}{\z@}%
                                     {-3.25ex\@plus -1ex \@minus -.2ex}%
                                     {1.5ex \@plus .2ex}%
                                     {\normalfont\bfseries}}
\makeatother

\let\non\nonumber

\def\lbldef#1#2{\expandafter\gdef\csname #1\endcsname {#2}}

\def\href#1#2{#2}

\newcommand{\beq}{\begin{equation}}
\newcommand{\eeq}{\end{equation}}

\def\bea{\begin{eqnarray}}
\def\eea{\end{eqnarray}}

\def\gs{\,\raise.15ex\hbox{/}\mkern-11.5mu G} 
\def\cs{\,\raise.15ex\hbox{/}\mkern-11.5mu C} 

\def\Pr{ { {\cal{P}}_{\cal{R}} }  }
\def\Prp{ { {\cal{P}}_{\cal{R^+}} }  }
\def\Prm{  {\cal{P}}_{ {\cal{R^-}} }  }
\def\cPpm{  {\cal{P}}_{ {\cal{R^\pm}} }  }
\def\cRp{ { { \cal{R}}^+}  }
\def\cRm{ { {\cal{R}}^-}  }
\def\cRpm{ { {\cal{R}}^\pm}  }
\def\cRmp{ { {\cal{R}}^\mp}  }
\def\cR{ {\cal{R}}  }

\def\Pb{\bar{P}}
\def\Gb{\bar{\Gamma}}
\def\cPb{  {\bar{\cal{P}}}  }
\def\cRb{  {\bar{\cal{R}}}  }
\def\eb{\bar{e}}
\def\gb{\bar{G}}
\def\gh{\hat{G}}
\def\fs4{ fuzzy $S^{4}$}
\def\fs3{ fuzzy $S^{3}$}
\def\fs2{ fuzzy $S^{2}$}
\def\mnc{$Mat_{N}(\mathbb{C})$}

\def\anst{${\cal A}_n(S^3)$ }
\def\angst{${\cal A}^{(g)}_n(S^3)$}

\providecommand{\openone}{\leavevmode\hbox{\small1\kern-3.8pt\normalsize1}}
\def\anstm{{\cal A}_n(S^3)}
\def\angstm{{\cal A}^{(g)}_n(S^3)}
\def\mncm{Mat_{N}(\mathbb{C})}

\begin{document}

\setcounter{page}{1}



\begin{titlepage}

\begin{center}

\hfill QMUL-PH-06-05\\
\hfill DAMTP-2006-36

\vskip 2 cm
{\Large \bf{A note on the M2-M5 brane system and fuzzy spheres}}\\

\vskip 1.25 cm {David S. Berman\footnote{email: D.S.Berman@qmul.ac.uk}}
\\
{\vskip 0.5cm
Queen Mary College, University of London,\\
Department of Physics,\\
Mile End Road,\\
London, E1 4NS, England.}\\
{\vskip 0.5cm
and}

\vskip 0.5 cm {Neil B. Copland\footnote{email: N.B.Copland@damtp.cam.ac.uk}}
\\
{\vskip 0.5cm
Department of Applied Mathematics and Theoretical Physics,\\
Centre for Mathematical Sciences,\\University of Cambridge,\\
Wilberforce Road,\\
Cambridge CB3 0WA,\\ England.}\\

\end{center}

\vskip 2 cm

\begin{abstract}
\baselineskip=18pt
This note covers various aspects of recent attempts to describe
membranes ending on fivebranes using fuzzy geometry. In particular, we
examine the Basu-Harvey equation and its relation to the Nahm equation
as well as the consequences of using a non-associative algebra for the
fuzzy three-sphere. This produces the tantalising result that the
fuzzy funnel solution corresponding to Q coincident membranes ending
on a five-brane has $Q^{3/2}$ degrees of freedom.

\end{abstract}

\end{titlepage}

\pagestyle{plain}

\baselineskip=19pt

\section{Introduction}

This paper will be concerned with describing how multiple membranes end on five-branes. From the five-brane world
volume point of view, membranes ending on a five-brane are described
by the self-dual string solution \cite{HLW} of the five-brane world
volume theory. The same system may also be described from the membrane world volume point of view
as a fuzzy three-funnel \cite{Basu:2004ed}.
This has since been generalised to include membranes ending on five-branes wrapped on
calibrated cycles \cite{bermancopland} and membranes stretching between several
five-branes\cite{norgardi}. 

To gain some intuition let us first consider the simpler case of D1 branes ending on D3 branes. This may be described in
two equivalent ways. One is as a monopole in the D3 brane world volume theory and the other is
as a fuzzy funnel solution to the Nahm equation,
\beq\label{Nahm}
\frac{dX^{i}}{d\sigma} - \frac{i}{2}\epsilon_{ijk}[X^{j},X^{k}]=0 \, ,
\eeq
which is the one half BPS equation in
the D1 brane world volume theory. The fuzzy funnel solution
\cite{Constable:1999ac} to (\ref{Nahm}) is essentially a fuzzy two-sphere whose radius
depends on the transverse distance from the brane with the two-sphere
opening up as it approaches the D3-brane.
The Nahm transformation is the shift between the D3-brane and D1-brane
points of view with the monopole describing the D1-brane from the
D3-brane perspective and the fuzzy funnel describing the D3-brane from
the D1-brane perspective \cite{Diaconescu,Tsimpis}.

The Basu-Harvey equation \cite{Basu:2004ed}
\beq\label{BaH}
\frac{dX^i}{d\sigma}+\frac{M_{11}^3}{8\pi\sqrt{2N}}\frac{1}{4!}\epsilon_{ijkl}[G_5,X^j,X^k,X^l]=0.
\eeq
is the M-theory analogue of the Nahm equation. It describes membranes
ending on a five-brane from the membrane world volume point of
view. Its solution is also a fuzzy funnel but for the membrane it is a fuzzy
three-sphere that opens up into the five-brane. This has been shown to
be consistent with what you would expect from comparing with the self-dual 
string solution of the five-brane.

Importantly though several issues remain unresolved for the Basu-Harvey equation. There is no
analogue of the Nahm transformation which would then map
solutions of the Basu-Harvey equation to the solutions of the
five-brane world volume theory and worryingly there is no supersymmetric
non-Abelian membrane theory that produces the Basu-Harvey equation as a one a half BPS
equation. 

Given its Nahm like role and the usual M-theory, string theory
relationship one would expect there to be a
{\it{reduction}} of the Basu-Harvey equation to the Nahm
equation. This paper will describe the relationship between the Basu-Harvey equation and the
Nahm equation. Effectively this will be a description of how to get fuzzy two-spheres from fuzzy three-spheres via a projection.

It has been noted that the appropriate algebra for a fuzzy
  three-sphere is non-associative \cite{Guralnik:2000pb, Ramgoolam:2001zx,
  Ramgoolam:2002wb}. 
We will examine the consequences of using such a non-associative algebra
 for the M2 brane geometry and find a tantalising result that the number of
 degrees of freedom scale as $N^{3/2}$ where N is the number of
 coincident membranes. One can interpret this the following way. The
 fuzzy three-sphere is naturally endowed with an ultraviolet cutoff. Summing all the spherical harmonics on the three-sphere up to
 the cutoff provides one with a finite number of degrees of freedom;
 the ratio of the size of the sphere to the cutoff being N
 dependent. It works out from the fuzzy sphere algebra that indeed
 the number of degrees then scale as $N^{3/2}$.

This is consistent with the view presented in \cite{Hofman:2001zt} which advocates
the presence of a non-associative algebra on five-branes in the presence
of background C flux. The self-dual string produces flux at the core
of the solution so it would be natural from this perspective to also expect
a non-associative algebra for the membrane five-brane
system. Alternative deformations of the five-brane geometry due to the
presence of flux have also been described in \cite{bergshoeffberman}.

Finally, it has been suggested \cite{baggerandlambert} that a non-associative
algebra may allow for a supersymmetric version of the coincident
membrane theory. That non-associativity is different from that explored
here but still it is interesting to consider novel algebras for the M2
brane system.

It should be noted that related work on fuzzy spheres has been studied
recently from various perspectives in \cite{Bhattacharyya:2005cd,Thomas:2006ac}. The paper is structured as follows. We first describe the basics of
the the fuzzy three-sphere and then examine its algebra and the
relation to the membrane five-brane system. Finally we relate the
Basu-Harvey equation to the Nahm equation through the projection from
fuzzy three-sphere to fuzzy two-sphere.

\section{The Fuzzy Three-Sphere and its Young Diagram Basis}

Fuzzy odd spheres are more complicated than the fuzzy even spheres\cite{Castelino:1997rv,Guralnik:2000pb, Ramgoolam:2001zx, Ramgoolam:2002wb}.
The construction of fuzzy odd spheres is derived by starting from the
fuzzy even sphere of one dimension higher and then applying a
projection. For the fuzzy three-sphere the starting point is thus the
fuzzy four-sphere. Given $V$, the four-dimensional spinor representation, on which the $Spin(5)$
$\Gamma$-matrices act, the fuzzy four-sphere co-ordinates are represented
by operators which act on $n$'th symmetrised tensor product of $V$. The co-ordinates $\gh^\mu$ are given by 
\beq
\gh^\mu=(\Gamma^\mu\otimes1\otimes1\otimes\ldots\otimes1+1\otimes\Gamma^\mu\otimes1\otimes\ldots\otimes1+\ldots+1\otimes\ldots\otimes1\otimes\Gamma^\mu)_{sym},
\eeq
(throughout $\mu,\nu,\ldots$ run from 1 to 5, $i,j,\dots$
from 1 to 4 and $a,b,\dots$ from 1 to 3). Some intuition can be gained
from the $n=1$ case where these are just the $Spin(5)$ Gamma matrices. $\rho_m(\Gamma^\mu)$ will
denote the
action of $\Gamma^\mu$ on the $m$'th factor of the tensor product and
$e_i$, $i=1,\ldots 4$, is the basis of $V$,
\beq
\rho_m(\Gamma^\mu)(e_{i_1}\otimes e_{i_2}\otimes\dots\otimes
e_{i_m}\otimes\ldots\otimes e_{i_n})=(e_{i_1}\otimes
e_{i_2}\otimes\dots\otimes (\Gamma^\mu_{j_mi_m}e_{j_m})\otimes\ldots\otimes
e_{i_n})
\eeq
and $P_n$ denotes symmetrisation so 
\beq
\gh^\mu=P_n\sum_m\rho_m(\Gamma^\mu)P_n.
\eeq
In their non-Abelian algebra the $\gh^\mu$ obey a version of the equation of a sphere,
$\sum_\mu \gh^\mu \gh^\mu=R^2\openone$, where $R$ is a function of
$n$. They also obey a higher Poisson bracket
equation\cite{SJ,SJ2} and commute with the generators of $SO(5)$, which are given
by $P_n\sum_m\rho_m(\Gamma^{[\mu}\Gamma^{\nu]})P_n$.

To get the fuzzy three-sphere we use the projection $P_\pm=\frac{1}{2}(1\pm\Gamma^5)$ to
decompose $V$ into $V_+$ and $V_-$, the positive and negative
chirality two-dimensional spinor representations of $SO(4)$. The
co-ordinates act in a reducible representation $\cR=\cRp\oplus\cRm$. To
get $\cRp$ we take the symmetrised tensor product of $\frac{n+1}{2}$ factors of
$V_+$ and $\frac{n-1}{2}$ factors of $V_-$, similarly $\cRm$ is the
symmetrised tensor product of $\frac{n-1}{2}$ factors of
$V_+$ and $\frac{n+1}{2}$ of $V_-$. $\cRp$ is the irreducible representation of
$SO(4)$ with $(2j_L,2j_R)=(\frac{(n+1)}{2},\frac{(n-1)}{2})$ and
$\cRm$ is the irreducible representation with $(2j_L,2j_R)=(\frac{(n-1)}{2},\frac{(n+1)}{2})$.  The projector
$\cPpm=\left(P_+^{\otimes(n\pm1)/2}\otimes P_-^{\otimes(n\mp1)/2}\right)_{sym}$ projects
the fuzzy four-sphere onto $\cRpm$. $\Pr=\Prp+\Prm$ and the co-ordinates of the fuzzy
three-sphere are given by
\beq
G^i=\Pr\gh^i\Pr.
\eeq
We can treat the $G^i$ as $N\times N$ matrices acting linearly on the
$N=\frac{(n+1)(n+3)}{2}$ dimensional representation
$\cR$. It can be shown that $\sum_i G^i G^i=\frac{(n+1)(n+3)}{2}\openone$ and
also\cite{Basu:2004ed} that
\beq\label{FS3}
G^i +\frac{1}{2(n+2)}\epsilon_{ijkl}G_5G^jG^kG^l =0,
\eeq
where $G_5=\Pr \gh^5 \Pr= \Prp-\Prm$. 

The space of $N\times N$ matrices acting on $N$-dimensional $\cR$,
\mnc, can be decomposed into representations of $SO(4)$, this is a basis of operators corresponding to
Young diagrams\cite{Ramgoolam:2001zx}. Young diagrams for $SO(4)$ have
a maximum of two rows (as
$\Gamma^1\Gamma^2\Gamma^3\Gamma^4=\Gamma^5$ so all products of more
than two $\Gamma$'s can be rewritten as products or two or less) and can be represented by the row lengths
$(r_1,r_2)$, where $r_2$ can be positive or negative. Each column is either
one or two boxes long and we represent these by factors of
$\rho_m(\Gamma)$ or $\rho_m(\Gamma\Gamma)$ respectively, each acting on
different factors of the tensor product. We suppress the indices on the
$\Gamma$'s. (They will be contracted with a tensor of appropriate
symmetry). If $r_2$ is positive, the
$\rho_m(\Gamma\Gamma)$ all act on $V_+$ factors, if it is negative
they all act on $V_-$ factors. If $r_1-r_2$ is divisible by two them
we are in $End(\cRpm)$ and half of the $\rho_m(\Gamma)$'s come with
$P_+$ projector ($\rho_m(\Gamma P_+)$) and half come with a $P_-$ to
make sure that we stay within $\cRpm$. If
$r_1-r_2$ is not divisible by two them
we are in $Hom(\cRpm,\cRmp)$ and $(r_1-r_2\pm1)/2$ of the
$\rho_m(\Gamma)$'s come with a
$P_+$ and $(r_1-r_2\mp1)/2$ with a $P_-$. 
For example a diagram in $Hom(\cRp,\cRm)$ with row lengths $(4,1)$ the operators would be
of the form 
\beq
\sum_{\vec{r},\vec{s}}\rho_{r_1}(\Gamma\Gamma P_+)\rho_{s_1}(\Gamma
P_-)\rho_{s_2}(\Gamma P_+)\rho_{s_3}(\Gamma P_+)
\eeq
and in $End(\cRm)$ with row lengths $(5,-3)$ they are of the form
\beq
\sum_{\vec{r},\vec{s}}\rho_{r_1}(\Gamma\Gamma P_-)\rho_{r_2}(\Gamma\Gamma P_-)\rho_{r_3}(\Gamma\Gamma P_-)\rho_{s_1}(\Gamma
P_+)\rho_{s_2}(\Gamma P_-).
\eeq
The sum over $\vec{r},\vec{s}$ is such that $r_1 \neq r_2\neq\ldots
s_1\neq s_2\neq\ldots$ and all indices run form 1 to $n$. The vector index on the $\Gamma$'s is contracted
with a tensor of appropriate Young diagram symmetry. The product of two
$\Gamma$'s ensures that we have anti-symmetry down columns and the
symmetrisation of $\cR$ ensures we have the correct symmetry along
rows. Group theoretical formulae for the number of independent
traceless tensors of this form for given row lengths are given in
\cite{Ramgoolam:2001zx}. Summing over allowed row lengths one finds
the number of independent operators is exactly $N^2$. (Diagrams with
both $\rho_r(\Gamma\Gamma P_+)$ and $\rho_r(\Gamma\Gamma P_-)$ factors
vanish, again see \cite{Ramgoolam:2001zx}.)

\section{Projection to Fuzzy Spherical Harmonics}

In the fuzzy three-sphere construction used in \cite{Basu:2004ed} the
co-ordinate matrices were taken to be in $Mat_N(\mathbb{C})$. As
mentioned there, this was not the only choice. We want an algebra
which reproduces the classical algebra of functions on
the $S^3$ in the large $N$ limit, \mnc\ does not. To see this let
$x^i=G^i/n$, so that $x^ix^i\sim O(1)$. Then
$x^{[i}x^{j]}\sim O(1)$ as well, and the anti-symmetric part of the
product of co-ordinates persists in the large $N$ limit. This
behaviour is caused by the second term of equation
(\ref{com}) below. For fuzzy even spheres the symmetry
over all $n$ tensor factors causes such a term to vanish. Here we have a
division between $V_+$ and $V_-$ factors. 

The projection which does give the correct limit is
detailed in \cite{Ramgoolam:2001zx}. It is a  projection onto operators
corresponding to Young diagrams with only one row (i.e. completely
symmetric), the algebra of these operators is called ${\cal A}_n(S^3)$\footnote{There is another possible
  algebra one can consider, the algebra generated by taking products of the
  $G^i$, \angst, however it does not have the correct large $N$ limit
  either. $\mncm\supset\angstm\supset\anstm$.}. We can extract the
terms corresponding to these operators
from the general sums given in \cite{Ramgoolam:2001zx}. One finds that
the number of operators surviving in $End(\cRpm)$ is $n(n+1)(n+2)/6$ and the
number in $Hom(\cRpm,\cRmp)$ is $(n+1)(n+2)(n+3)/6$. The other
condition of the projection is that matrices should act in the same
manner on $\cRp$ and $\cRm$ so we sum each state of 
$End(\cRp)$ with the corresponding one form $End(\cRm)$ and so on,
giving the total number of degrees of freedom as
\beq
D=(n+1)(n+2)(2n+3)/6.
\eeq

However, the algebra ${\cal A}_n(S^3)$ does not close under
multiplication, and we have to project back onto ${\cal A}_n(S^3)$
after multiplying. We denote the projected product by $A\bullet B$ or $(AB)_+$ for
$A,B\in {\cal A}_n(S^3)$. This new product is non-associative. (In the
large $n$ limit one recovers associativity \cite{Ramgoolam:2001zx,Papageorgakis:2006ed}). The
mechanism for non-associativity can be easily seen for the simple $n=1$ example
where 
\beq
(\Gamma^1\bullet\Gamma^1)\bullet\Gamma^2=\openone\bullet\Gamma^2=\Gamma^2\neq0=\Gamma^1\bullet0=\Gamma^1\bullet(\Gamma^1\bullet\Gamma^2).
\eeq
Generalising to higher $n$ we must be careful to decompose into the
Young diagram basis and keep only the proscribed set. In particular we
should remember to keep traces of tensor operators that correspond to
smaller symmetric diagrams. Firstly taking the projected product of
two of the co-ordinates we get
\beq
G^i\bullet G^j=\Prp\left[\sum_r\rho_r(\delta^{ij}P_+)+\sum_{r\neq
  s}\rho_r(\Gamma^{(i}P_-)\rho_s(\Gamma^{j)}P_+)\right]\Prp\;+\;
  (+\leftrightarrow-).
\eeq
We then use this to multiply a third co-ordinate giving
\bea
(G^i\bullet G^j)G^k&=&\Prp\left[\sum_{r\neq s\neq
    t}\rho_r(\Gamma^{(i}P_-)\rho_s(\Gamma^{j)}P_+)\rho_t(\Gamma^{k}P_-)+\sum_{r\neq s}\rho_r(\delta^{ij}P_+)\rho_s(\Gamma^kP_-)\right.\non\\
 &&+\sum_r\rho_r(\delta^{ij}\Gamma^kP_-)+\frac{1}{2}\sum_{r\neq s}\rho_r(\Gamma^{i}P_-)\rho_s(\Gamma^j\Gamma^kP_-)+\non\\
 &&\left.\frac{1}{2}\sum_{r\neq s}\rho_r(\Gamma^{j}P_-)\rho_s(\Gamma^i\Gamma^kP_-)\right]\Prm\;+
\;  (+\leftrightarrow-).
\eea
Now we must be careful to keep the traces of the last term when we
project. The traceless combination is
\beq
\frac{1}{2}\sum_{r\neq
  s}\left(\rho_r(\Gamma^{i}P_-)\rho_s(\Gamma^j\Gamma^kP_-)+\rho_r(\Gamma^{j}P_-)\rho_s(\Gamma^i\Gamma^kP_-)\right)-\frac{n-1}{6}\left(\delta^{ij}G^k+\delta^{ik}G^j+\delta^{jk}G^i\right),
\eeq
as can be checked by contracting with a delta function. The trace parts
correspond to $(r_1,r_2)=(1,0)$ diagrams so we keep them. Projecting
the second product therefore gives
\bea
(G^i\bullet G^j)\bullet G^k&=&\Prp\left[\sum_{r\neq
  s\neq
  t}\rho_r(\Gamma^{(i}P_-)\rho_s(\Gamma^{j}P_+)\rho_t(\Gamma^{k)}P_-)+\right.\non\\ &&\left.\frac{2n+1}{3}\delta^{ij}G^k+\frac{n-1}{6}\left(\delta^{ik}G^j+\delta^{jk}G^i\right)\right]\Prm\;+\;
  (+\leftrightarrow-).
\eea
A similar calculation can be done for $G^i\bullet (G^j\bullet G^k)$
and the non associativity can be expressed by
\beq
(G^i\bullet G^j)\bullet G^k-G^i\bullet(G^j\bullet G^k)=\frac{n+1}{2}(\delta^{ij}G^k-\delta^{jk}G^i),
\eeq
where we have a different overall factor to
\cite{Papageorgakis:2006ed} due to making only keeping the
symmetric part after the first projection. 

One can ask if the Basu-Harvey equation still holds for this projection, and we
find that it does not. As one might expect the antisymmetric bracket
vanishes when we project onto symmetric representations. The co-ordinates
$G^i$ are contained in ${\cal A}_n(S^3)$. The commutator of two
of these co-ordinate matrices is given by
\bea
[G^{i}, G^{j}]\cPpm&=&2\sum_{r}\rho_r(\Gamma^{ij}P_\pm)\cPpm +\sum_{r\neq
  s}2\rho_r(\Gamma^{[i} P_\mp)\rho_s(\Gamma^{j]} P_\pm)\cPpm\label{com}\\
&=&2\sum_{r}\rho_r(\Gamma^{ij}P_\pm)\cPpm-\frac{n+1}{2}\sum_{r}\rho_r(\Gamma^{ij}P_{\mp})\cPpm\non\\
&&+\frac{n-1}{2}\sum_{r}\rho_r(\Gamma^{ij}P_{\pm})\cPpm
\eea
where in the second line we have written everything in terms of the
Young diagram basis of \cite{Ramgoolam:2001zx}. We see that written in
this basis every term contains a product of two $\Gamma$'s acting on
the same tensor product factor. This corresponds to Young diagrams with $r_2\neq0$ so
after applying the projection there are no surviving terms. 

Since the two bracket vanishes, the anti-symmetric four bracket also
does. It seems therefore the algebra ${\cal A}_n(S^3)$ is not compatible
with the Basu-Harvey equation. However the algebra ${\cal
  A}_n(S^3)$ has a tantalising property. If we take the solutions to
the Basu-Harvey equation
\beq
X^i(\sigma)=\frac{i\sqrt{2\pi}}{M_{11}^\frac{3}{2}}\frac{1}{\sqrt{\sigma}}G^i
\eeq
and look at the physical radius,
\beq\label{traces}
R=\sqrt{\left|\frac{Tr \sum(X^i)^2}{Tr \openone}\right|}
\eeq
we get
\beq
\sigma=\frac{2\pi N}{M_{11}^3R^2}\, .
\eeq
(Notice that in (\ref{traces}) we no longer have a matrix trace with
our modified algebra, however because $\sum_i X^iX^i$ is proportional
to the identity in the algebra, and trace of the identity is precisely
what we divide by to obtain the physical radius, the form of the trace
is unimportant here). If we still identify $N$ with $Q$, the number of membranes, then we reproduce the profile expected from the self-dual string.

However, now the number of degrees of freedom in the co-ordinates is no longer $N^2$, but $D=(n+1)(n+2)(2n+3)/6\sim n^3$ so that for large $N$ (thus $n$) we have that
\beq
D\sim Q^\frac{3}{2}
\eeq
exactly as expected for $Q$ coincident membranes in the large $Q$
limit. This is interpreted as the result of the fuzzy three-sphere
being endowed with an ultraviolet cutoff and the scaling of the cutoff to the size
of sphere depends in the right way on $Q$ to give the correct number
of degrees of freedom. This means that one can interpret the $Q^{3/2}$
degrees of freedom corresponding to the non-Abelian membrane theory as coming
from modes on the fuzzy sphere.

This is encouraging but we must reconcile the non-associative
projection and the four-bracket. One possibility is that the projection does not act inside
the bracket, which is thought of as an operator $[G_5,\cdot,\cdot, \cdot]:({\cal
  A}_n(S^3))^3\rightarrow{\cal A}_n(S^3)$. In this case obviously
$X^i$ will still provide
a solution. (See the following section for how something similar
happens for dimensional reduction.)

\section{Reducing the Fuzzy Three-Sphere to the Fuzzy Two-Sphere}\label{dimred}

Reducing the fuzzy three-sphere to the fuzzy two-sphere comes down to
finding a projection on the fuzzy three-sphere so that three of the fuzzy three-sphere
matrices obey the fuzzy two-sphere algebra.

We begin at n=1 and look for a projector $\bar{P}$ such that it
commutes with $\Gamma^a,\, a=1,2,3$ and
$[\Gamma^a,\Gamma^b]\Pb=2i\epsilon_{abc}\Gamma^c\Pb$. Looking ahead to
the Basu-Harvey equation, we will also require that $\Pb\Gamma^4\Pb=0$ to satisfy the fuzzy three-sphere equation. We use a
basis given by Appendix \ref{conv} and assume that the projector is
made of
$2\times 2$ blocks proportional to the identity, i.e. constructed
from ${\openone,\Gamma^4,\Gamma^5,\Gamma^{45}}$. The solution is given
by 

\beq
\bar{P}=\frac{1}{2}(1+i\Gamma^4\Gamma^5).
\eeq
One can then clearly see $\Pb^2=\Pb$, $\Pb$
commutes with $\Gamma^a$ and $\Pb\Gamma^4\Pb=\Pb\Gamma^5\Pb=0$.

Defining $\Gb^\mu=\Pb\Gamma^\mu\Pb$ we now have
\beq
[\Gb^a,\Gb^b]=2i\epsilon_{abc}\Gb^c,
\eeq
in other words, when restricted to the subspace which $\Pb$ projects onto,
the $\Gamma^a$ form an $SU(2)$ algebra. This is easy to see when we
choose $\{\eb_1=\frac{1}{2}(e_1+ie_3), \eb_2=\frac{1}{2}(e_2+ie_4)\}$
as a basis for $\Pb V$. Then
\beq\label{gsigma}
 \Gb^a(\eb_i)\equiv\sigma^a_{ji}\eb_j
\eeq
 where
$\sigma^a$ are the Pauli matrices.

We can generalise to any $n$ by
introducing $\cPb=\Pb^{\otimes n}$ which projects onto
$\cRb=(\Pb V)^{\otimes n}$. We denote the original fuzzy four-sphere
matrices, which act on $V^{\otimes n}$, by $\gh^\mu$. Then we
set
\beq
\gb^\mu=\cPb \gh^\mu\cPb= \cPb \sum_r\rho_r(\Gamma^\mu)\cPb
\eeq
which is just saying that we have restricted $\gh^\mu$ to $\cRb$. Now
because of (\ref{gsigma}) we will recover the construction of the fuzzy
two-sphere given in Appendix B of \cite{SJ2}. Thus one can check that

\bea \label{projS2}
[\gb^a,\gb^b]=2i\epsilon_{abc}\gb^c, \\
\gb^a\gb^a=n(n+2)\openone.
\eea

Notice that if we define the co-ordinate matrices of the fuzzy three-sphere with a projector onto $\cR$ either
 side of them  ($G^{i}=\Pr \gh^{i} \Pr$) then $\gb^{a}\neq
 \cPb G^{a}\cPb$. However $\gb^{a}\propto\cPb G^{a}\cPb$ 
 because $\Pb P_\pm\Pb=\Pb/2$. Thus the constant of proportionality has a
 power of $2$. It also has combinatoric factor dependent as there are
 ${(n-1) \choose (n-1)/2}$ ways of choosing which tensor
 product factors are acted upon by the $ \frac{n-1}{2}$ $P_+$'s in $\Pr$ which
 do not act on the same factor as the $\Gamma$. In $\cPb$ we just have
 $n-1$ $\Pb$ factors so only one choice. Hence the constant of
 proportionality is given by
\beq\label{factor}
\gb^{a}={(n-1) \choose (n-1)/2}^{-1}2^{n-1}\cPb G^{a}\cPb.
\eeq
 The projectors are there to indicate that we project on to $\cR$ or $\cRb$, so this is not a
 problem; we can think of acting with the same original $\gh^\mu$
 of the fuzzy $S^4$ in both
 cases before we project back to the representation we are dealing
 with using $\Pr$ or $\cPb$. 

To make sure we can project from the fuzzy two-sphere to the fuzzy
 three-sphere what we should check is that for any state,
$\Psi$,  in $\cRb$ we can find state in $\cR$ such that $\cPb$ projects it onto $\Psi$. Indeed we can find many such
states. Similarly for any operator on these states in the fuzzy
 two-sphere we can find operators in the full fuzzy three-sphere algebra,
 \mnc, that project on to it. In fact if we restrict ourselves to the
 non-associative algebra \anst there is a unique operator that
 projects onto each operator in the fuzzy two-sphere, up to addition of
 operators in the kernel of $\cPb$. A general operator of the
 form
\beq
\cPb\sum_{\overrightarrow{r}\neq\overrightarrow{s}\neq\overrightarrow{t}}\rho_{r_1}(\Gamma^1)\rho_{r_2}(\Gamma^1)\dots\rho_{r_i}(\Gamma^1)\rho_{s_1}(\Gamma^2)\dots\rho_{s_j}(\Gamma^2)\rho_{t_1}(\Gamma^3)\dots\rho_{t_k}(\Gamma^3)\cPb
\eeq
is proportional to that obtained by projecting
\bea
\Pr\sum_{\overrightarrow{r}\neq\overrightarrow{s}\neq\overrightarrow{t}}\rho_{r_1}(\Gamma^1P_\pm)\rho_{r_2}(\Gamma^1P_\pm)\dots\rho_{r_i}(\Gamma^1P_\pm)\rho_{s_1}(\Gamma^2P_\pm)\dots\rho_{s_j}(\Gamma^2P_\pm)\nonumber\\
\rho_{t_1}(\Gamma^3P_\pm)\dots\rho_{t_k}(\Gamma^3P_\pm)\Pr
\qquad +\qquad (+\leftrightarrow -).
\eea
where the signs of each $P_\pm$ are chosen to alternate from right to
 left. In both the fuzzy three-sphere and the fuzzy two-sphere we are still
 effectively using
 the co-ordinate matrices of the fuzzy four-sphere, but we are restricting
 to a much reduced set of states.

Notice also if we plug the our projected fuzzy three-sphere matrices
straight back into the fuzzy three-sphere equation (\ref{FS3}) then it
is {\it not} satisfied due to the vanishing of $\gb^{4}$. This should
be compared with taking the classical version of the three-sphere
equation and reducing to two-sphere (or any other sphere reduction) where the Nambu bracket is replaced by a Poisson
bracket: if we reduce to a sub-sphere of lower degree, say by fixing one of the co-ordinates, then the equation will
not be satisfied. This occurs when the derivatives in the Poisson bracket act on the constant co-ordinate. However if we set the
co-ordinate to its fixed value \textit{after} evaluating the
derivatives then the higher sphere equation will still be satisfied. 

For the Nambu bracket there are no derivatives and the information is in the anti-commutation properties of the
matrices. Hence we should make our projection \textit{after}
evaluating the Nambu-bracket in (\ref{FS3}). In our case we see that our
projected sphere satisfies the higher sphere equation trivially.
The main change is that both terms vanish for $i=4$, a necessity to
obey the Basu-Harvey equation below.

We can also ask how the $SU(2)$ of the fuzzy two-sphere fits inside the
$SU(2)\times SU(2)=SO(4)$ of the fuzzy three-sphere. The $SO(4)$ has the six generators 
\beq
G^{ij}=\Pr\sum_r\rho_r(\Gamma^{ij})\Pr.
\eeq

>From these we can construct two orthogonal $SU(2)$'s by 
\beq
\Sigma^{ij}_{L/R}=-i(G^{ij}\mp\epsilon_{ijkl}G^{kl})=-i\Pr\sum_r\rho_r(\Gamma^{ij}P_\pm)\Pr.
\eeq
The $\Sigma^{ij}_L$ and the $\Sigma^{ij}_R$ form $SU(2)$ algebras among themselves and commute with each other, we call these $SU(2)$'s $SU(2)_L$
and $SU(2)_R$ respectively. A general state in $\cRpm$ of the the fuzzy
$S^3$ is given by 
\beq
\left((e_1)^{\otimes p}\otimes (e_2)^{\otimes \frac{n\pm1}{2}-p}\otimes
(e_3)^{\otimes q}\otimes (e_4)^{\otimes \frac{n\mp1}{2}-q}\right)_{sym}
\eeq
where $e_i$ is the basis of $V$. If we label the generators of the two $SU(2)$'s by
$\sigma^a_{L/R}=1/2\epsilon_{abc}\Sigma^{bc}_{L/R}$ then these basis states are
eigenstates of $\sigma^3_{L/R}$, with eigenvalues $2m_{L/R}$. We can then
apply $\cPb$ to these states and examine the eigenvalues of $\gb^3$ on
these new $\cRb$ states,
$2\bar{m}$ say. Then we find that $\bar{m}=m_L+m_R$. Since adding the
size of the reps of $SU(2)_L$ and $SU(2)_R$ in $\cRpm$ gives
$(n\pm1)/2+(n\mp1)/2=n$, which is the size of the rep of the fuzzy three-sphere $SU(2)$, we see that we are effectively taking the sum of
$SU(2)_L$ and $SU(2)_R$.

\section{Disappearance of Non-associativity after Dimensional
  Reduction}

Given the non-associative nature of the fuzzy three-sphere algebra, it is
natural to ask how it gives rise to the associative algebra of the
fuzzy two-sphere. In the fuzzy two-sphere there exist only matrices
corresponding to symmetric Young diagrams. This is because the su(2) algebra
$\sigma^i \sigma^j =\delta^{ij}\openone+\epsilon^{ijk}\sigma^k$ implies
that an anti-symmetrised product of $\sigma$'s can be written as a single $\sigma$.

We can perform a check that the product in the fuzzy two-sphere is
associative,
\beq
\gb^a\gb^b\cPb=\left(\sum_r\rho_r(\delta^{ab}\openone+i\epsilon_{abd}\Gamma^d)+\sum_{r\neq
  s}\rho_r(\Gamma^a)\rho_s(\Gamma^b)\right)\cPb
\eeq
so that
\bea
(\gb^a\gb^b)\gb^c\cPb&=&\left(\sum_r\rho_r(\delta^{ab}\Gamma^c+i\epsilon_{abd}\Gamma^d\Gamma^c)+\sum_{r\neq
  s}\rho_r(\Gamma^a\Gamma^c)\rho_s(\Gamma^b)\right)\non\\
&&\left.+\sum_{r\neq
  s}\rho_r(\Gamma^a)\rho_s(\Gamma^b\Gamma^c)+\sum_{r\neq
  s}\rho_r(\delta^{ab}+i\epsilon_{abd}\Gamma^d)\rho_s(\Gamma^c)+\sum_{r\neq s\neq t}\rho_r(\Gamma^a)\rho_s(\Gamma^b)\rho_t(\Gamma^c)\right)\cPb\non\\
&=&\left(\sum_r\rho_r(\delta^{ab}\Gamma^c+i\epsilon_{abd}(\delta^{cd}\openone+i\epsilon_{dce}\Gamma^e))+\sum_{r\neq
  s}\rho_r(\delta^{ac}\openone+i\epsilon_{acd}\Gamma^d)\rho_s(\Gamma^b)\right.\non\\
&&\left.+\sum_{r\neq
  s}\rho_r(\Gamma^a)\rho_s(\delta^{bc}\openone+i\epsilon_{bcd}\Gamma^d)+\sum_{r\neq
  s}\rho_r(\delta^{ab}+i\epsilon_{abd}\Gamma^d)\rho_s(\Gamma^c)\right.\non\\
&&\left.+\sum_{r\neq s\neq
  t}\rho_r(\Gamma^a)\rho_s(\Gamma^b)\rho_t(\Gamma^c)\right)\cPb\non
\eea
\bea
\qquad\qquad\quad&=&\Biggl( in\epsilon_{abc}+n\delta^{ab}\gb^c+(n-2)\delta^{ac}\gb^b+n\delta^{bc}\gb^a
\non\\&&\left.+i\epsilon_{acd}\sum_{r\neq s}\rho_r(\Gamma^d)\rho_s(\Gamma^b)
+i\epsilon_{bcd}\sum_{r\neq s}\rho_r(\Gamma^d)\rho_s(\Gamma^a)
+i\epsilon_{abd}\sum_{r\neq s}\rho_r(\Gamma^d)\rho_s(\Gamma^c)\right.\non\\
&&\left.+
\sum_{r\neq s\neq
  t}\rho_r(\Gamma^a)\rho_s(\Gamma^b)\rho_t(\Gamma^c)\right)\cPb
\eea
and similarly 
\bea
(\gb^a\gb^b)\gb^c\cPb&=&\Biggl( in\epsilon_{abc}+n\delta^{ab}\gb^c+(n-2)\delta^{ac}\gb^b+n\delta^{bc}\gb^a\non\\&&\left.+i\epsilon_{acd}\sum_{r\neq s}\rho_r(\Gamma^d)\rho_s(\Gamma^b)
+i\epsilon_{bcd}\sum_{r\neq s}\rho_r(\Gamma^d)\rho_s(\Gamma^a)
+i\epsilon_{abd}\sum_{r\neq s}\rho_r(\Gamma^d)\rho_s(\Gamma^c)\right.\non\\
&&\left.+
\sum_{r\neq s\neq
  t}\rho_r(\Gamma^a)\rho_s(\Gamma^b)\rho_t(\Gamma^c)\right)\cPb
\eea
so we have an associative product as expected. All terms can be written in terms
of symmetric operators as in the final line.

\section{Reducing the Basu-Harvey Equation to the Nahm Equation}

Consider the fuzzy three-sphere equation (\ref{FS3}) but replace the $G^i$  
by unknowns $\tilde{G}^i$ for which we must solve. Let us reduce to the  
fuzzy two-sphere so that the equation acts on $\cRb$,
\beq
\cPb \left(\tilde{G}^i  
+\frac{1}{2(n+2)}\epsilon_{ijkl}G_5\tilde{G}^j\tilde{G}^k\tilde{G}^l\right)\cPb =0.
\eeq
We let $\tilde{G}^4=G^4/c$ where $c$ is the factor arising from  
projection we saw previously in equation (\ref{factor}), but fix the  
$\tilde{G^a}$ to be matrices in the algebra of the fuzzy two-sphere
($\tilde{G^a}=\cPb\tilde{G^a}\cPb$. Then, because $\cPb G^4\cPb=0$ and  
$\cPb G^5G^4\cPb=inc\cPb$, the $\tilde{G}^a$ must obey
\beq
\cPb \left(\tilde{G}^a +\frac{in}{2(n+2)}\epsilon_{abc}\tilde{G}^b\tilde{G}^c\right)\cPb =0.
\eeq
In  the large $n$ limit this is the statement that the $\tilde{G}^a$  
must obey the fuzzy two-sphere $SU(2)$ algebra
\beq
[\tilde{G}^a,\tilde{G}^b]\cPb=2i\epsilon_{abc}\tilde{G}^c\cPb
\eeq
which of course the $\gb^a$ obey(\ref{projS2}). Having to take the  
large $n$ limit is expected because the fuzziness will make picking out  
a cross-section of the sphere difficult at small $n$.

We can now follow a similar procedure for the Basu-Harvey equation. We  
must take into account the additional anti-symmetrisation of the  
4-bracket and also use $R_{11}=M_{11}^{-3}\alpha'^{-1}$. We set  
$X^4=\frac{32\pi R_{11} G^4}{3c}$ and restrict the equation and the  
$X^a$ to the fuzzy two-sphere. Then we get
\beq
\left(\frac{dX^a}{d\sigma}+\frac{in}{\alpha'\sqrt{2N}}\epsilon_{abc}X^b 
X^c\right)\cPb=0.
\eeq
(There is also the case when the free index is $4$, here both terms  
vanish. This is required if we are to get the right $\sigma$  
dependence.) Rearranging, again in the large $n$ limit, we get
\beq
\frac{dX^{a}}{d\sigma} +  
\frac{i}{2\alpha'}\epsilon_{abc}[X^{b},X^{c}]=0 \, ,
\eeq
which is (\ref{Nahm}) but with the $X^a$ scaled by $\alpha'$ so that  
they have dimensions of length. It has solution
\beq
X^{a}=\frac{\alpha'\gb^{a}}{2\sigma} .
\eeq
The appearance of $R_{11}$ is expected as the length scale in  
11-dimensions is $1/M_{11}$ and in string theory it is  
$\sqrt{\alpha'}$. Changing from a three bracket for the Basu-Harvey  
equation to a two bracket for the Nahm produces a  
$M_{11}^{-3}\alpha'^{-1}=R_{11}$. Note, here we have considered either the  
case where the solutions before projection act in \mnc, or the case  
where they are in \anst but the projection does not act within the  
anti-symmetrised product.

\section{Discussion}
There is still much to be learnt about the M2-M5 brane system. The
tantalising appearance of $N^{3/2}$ for the non-associative algebra
suggests there may be something in the algebra of the fuzzy three-sphere that is relevant for the M2-M5 system even though the
projection discussed here appears not to be consistent with the Basu-Harvey
equation. The relation to the Nahm equation has been clarified but 
until the many unresolved questions are answered concerning the
supersymmetric non-Abelian membrane theory this discussion remains conjectural.

\section*{Acknowledgements}
We wish to thank Anirban Basu, Neil Lambert, Costis Papageorgakis 
and Sanjaye Ramgoolam for discussions. DSB is supported by EPSRC grant
GR/R75373/02 and would like to thank DAMTP and Clare Hall college
Cambridge for continued support. NBC is supported by a PPARC
studentship. This work was in part supported by the EC Marie Curie
Research Training Network, MRTN-CT-2004-512194.

\appendix
\section{Conventions}\label{conv}

We use the following basis for the $Spin(4)$ $\Gamma$ matrices:
\begin{equation}\label{gamma-convention}
\Gamma^{i} = \left(\begin{matrix} 0 && \sigma^i \\
\bar\sigma^i && 0 \end{matrix}\right),\ \Gamma^5 =
\left(\begin{matrix} \openone_{2\times2} && 0 \\ 0 &&
-\openone_{2\times2} \end{matrix}\right).
\end{equation}
where
\begin{equation}\label{sigma-sigma-bar}
\sigma^i = (-i\vec{\sigma}_{Pauli}, \openone_{2\times2}),\
\bar\sigma^i = ( i\vec{\sigma}_{Pauli}, \openone_{2\times2})
\end{equation}
with  $\vec{\sigma}_{Pauli}$ being the standard Pauli sigma
matrices:
\begin{equation}
\sigma^1 = \left(\begin{matrix}  0 && 1 \\ 1 && 0
\end{matrix}\right),\ \sigma^2 = \left(\begin{matrix}  0 && -i \\ i &&
0 \end{matrix}\right),\ \sigma^3 = \left(\begin{matrix}  1 && 0 \\
0 && -1 \end{matrix}\right) .
\end{equation}
Thus
$\Gamma^5=\Gamma^1\Gamma^2\Gamma^3\Gamma^4=\frac{1}{4!}\epsilon_{ijkl}
\Gamma^i\Gamma^j\Gamma^k\Gamma^l$.



\end{document}